\documentclass[twocolumn,prd,showpacs,
floatfix,preprintnumbers,nofootinbib]{revtex4}

\usepackage{epsfig}
\usepackage{amsmath,bm,color}

\begin{document}

\title{Neutrino flavor pendulum in both mass hierarchies}

\author{Georg Raffelt and David de Sousa Seixas}
\affiliation{Max-Planck-Institut f\"ur Physik
(Werner-Heisenberg-Institut), F\"ohringer Ring 6, 80805 M\"unchen,
Germany}

\date{29 July 2013, updated 22 August 2013}


\begin{abstract}
We construct a simple example for self-induced flavor conversion in
dense neutrino gases showing new solutions that violate the
symmetries of initial conditions. Our system consists of two opposite
momentum modes 1 and 2, each initially occupied with equal densities
of $\nu_e$ and $\bar\nu_e$. Restricting solutions to symmetry under
$1\leftrightarrow 2$ allows for the usual bimodal instability
(``flavor pendulum'') in the inverted neutrino mass hierarchy (IH)
and stability (no self-induced flavor conversion) in the normal
hierarchy (NH). Lifting this symmetry restriction allows for a second
pendulumlike solution that occurs in NH where the modes 1 and 2 swing
in opposite directions in flavor space. Any small deviation from 1--2
symmetry in the initial condition triggers the new instability in NH.
This effect corresponds to the recently identified multi-azimuth
angle (MAA) instability of supernova neutrino fluxes. Both cases show
explicitly that solutions of the equations of collective flavor
oscillations need not inherit the symmetries of initial conditions,
although this has been universally assumed.
\end{abstract}

\preprint{MPP-2013-214}


\maketitle

\section{Introduction}
\label{sec:introduction}

Neutrino-neutrino refraction can produce unexpected flavor
oscillation phenomena in the form of self-induced flavor conversion
in dense neutrino gases \cite{Pantaleone:1992eq, Sigl:1992fn,
Samuel:1993uw, Kostelecky:1993dm, Samuel:1995ri, Pantaleone:1998xi,
Dolgov:2002ab, Friedland:2003dv, Sawyer:2005jk, Duan:2005cp,
Duan:2006an, Hannestad:2006nj, Friedland:2006ke, Raffelt:2007yz,
Raffelt:2007cb, Duan:2007fw, Fogli:2007bk, Fogli:2008pt,
Sawyer:2008zs, Raffelt:2008hr, EstebanPretel:2008ni,
Dasgupta:2009mg, Sawyer:2010jk, Duan:2010bf, Raffelt:2010za,
Banerjee:2011fj, Sarikas:2011am, Raffelt:2011yb, Pehlivan:2011hp,
Dvornikov:2011dv, Chakraborty:2011gd, Saviano:2012yh,
Mirizzi:2012wp, Sarikas:2012ad, Cherry:2012zw, Sarikas:2012vb,
deGouvea:2012hg, Volpe:2013uxl, Cherry:2013mv}, for a review see
Ref.~\cite{Duan:2010bg}. One major complication is caused by the
current-current nature of low-energy weak interactions. It implies
that the refractive effect felt by a neutrino with velocity vector
${\vec v}$ caused by other neutrinos with velocity ${\vec v}'$ is
proportional to $1-{\vec v}\cdot{\vec v}'$. If we study the flavor
evolution of a homogeneous and isotropic neutrino gas (early
universe) a seemingly obvious simplification is to integrate out the
velocities, i.e., to drop the ${\vec v}\cdot{\vec v}'$ term. The
underlying assumption is that the solution of the equations of
motion (EoM) will be homogeneous and isotropic as well. Moreover, it
was tacitly assumed that unavoidable small deviations from these
symmetries in the initial system will remain small in the solution
as well. In the supernova (SN) context, the assumption of axially
symmetric neutrino emission was used to integrate out the azimuth
angle of neutrino propagation.

These universal assumptions go back to the earliest papers on the
subject, but are nevertheless unjustified. Based on a linearized
stability analysis in the context of axially symmetric SN neutrino
streams, we have recently found a new class of solutions
\cite{Raffelt:2013rqa}. They spontaneously break axial symmetry and
lead to azimuth-angle dependent flavor conversion effects. For simple
neutrino spectra, the traditional bimodal instability occurs in IH,
whereas the new multi-azimuth angle (MAA) instability occurs in NH.
Meanwhile, numerical studies by other authors confirm the existence
of this effect and its strong impact on collective flavor
oscillations \cite{Mirizzi:2013rla}.

This change of paradigm calls into question numerous results in the
previous literature. Any solution found under an uncontrolled
symmetry assumption may have missed the dominant effect.

The mode analysis that has led to the discovery of the MAA
instability is not complicated. The stability condition follows from
an eigenvalue equation derived from the linearized EoMs. However, one
finds stable or unstable conditions based on mathematical criteria
that are not necessarily physically transparent or intuitive. We here
provide a simple physical explanation in the simplest possible
example that shows a new instability after a symmetry assumption has
been relaxed.

Our system consists of two opposite moving beams 1 and 2, each
consisting of equal fluxes of $\nu_e$ and $\bar\nu_e$, all with the
same energy. The modes 1 and 2 are taken to be identically prepared
and therefore the solution should be symmetric under the exchange
$1\leftrightarrow2$. We consider two-flavor oscillations with a small
mixing angle.  In IH one then finds the well-known bimodal
instability \cite{Kostelecky:1993dm, Samuel:1995ri, Duan:2005cp} that
leads to pendulumlike oscillations \cite{Hannestad:2006nj} between,
say, $\nu_e\bar\nu_e\leftrightarrow\nu_\tau\bar\nu_\tau$. In NH the
system is stable and the motion consists of small-amplitude harmonic
oscillations.

After lifting the 1--2 symmetry assumption, the smallest disturbance
could trigger a new instability. Indeed, one finds a new pendulumlike
solution, now in NH, with the same flavor conversion effect. In IH,
the new solution is stable. Unless the system is prepared {\it
exactly\/} symmetric under $1\leftrightarrow2$ exchange, a
pendulumlike solution appears in both hierarchies.

We describe the flavor content with polarization vectors ${\bf
P}_{1,2}$ for neutrinos and $\bar{\bf P}_{1,2}$ for
anti-neutrinos.\footnote{We denote vectors in flavor space in
bold-face characters, whereas vectors in coordinate space are
denoted with an arrow.}\footnote{Much of the literature uses the
flavor-isospin convention where the polarization vectors describing
neutrinos and anti-neutrinos of the same flavor point in opposite
directions. In our convention they point in the same direction,
allowing for a more straightforward visualization of self-induced
flavor conversion, although in the equations we need to distinguish
explicitly between $\nu$ and $\bar\nu$.} In the limit of large
neutrino-neutrino refraction, the traditional solution consists of
all four polarization vectors sticking closely together and swinging
in flavor space like an inverted plane pendulum. The new solution
consists of ${\bf P}_{1}$ and $\bar{\bf P}_{1}$ to stick together as
well as ${\bf P}_{2}$ and $\bar{\bf P}_{2}$, but these two pairs now
pendulate in opposite directions in a single plane. This solution
can arise only if initially the four vectors are not exactly aligned
with each other.

It is actually simple to graphically understand these solutions and
we begin, in Sec.~\ref{sec:pics}, with a pictorial explanation. We
then turn in Sec.~\ref{sec:oscillators} to a mathematical
description, leading to two coupled anharmonic oscillators. We
conclude in Sec.~\ref{sec:conclusion}.

\section{Pictorial explanation}
\label{sec:pics}

To make contact with the earlier literature on the flavor pendulum
\cite{Hannestad:2006nj} we use the same notation and describe the
flavor content of the neutrino modes with polarization vectors in
flavor space ${\bf P}$ and $\bar{\bf P}$, where overbarred
quantities refer to antiparticles. If we consider $N$ momentum modes
with momenta ${\vec p}_i$ ($i=1,\ldots,N$), the vacuum oscillation
frequency of each mode is $\omega_i=\Delta m^2/2E_i$ and the
velocity vectors are ${\vec v}_i={\vec p}_i/E_i$. The oscillation
equations for neutrino and antineutrino mode $j$ is
\begin{eqnarray}\label{eq:eom1}
 \partial_t{\bf P}_j&=&\left[+\omega_j{\bf B}
 +\mu\sum_{i=1}^N\left({\bf P}_i-\bar{\bf P}_i\right)(1-{\vec v}_i\cdot{\vec v}_j)\right]\times{\bf P}_j\,,
 \nonumber\\
 \partial_t\bar{\bf P}_j&=&\left[-\omega_j{\bf B}
 +\mu\sum_{i=1}^N\left({\bf P}_i-\bar{\bf P}_i\right)(1-{\vec v}_i\cdot{\vec v}_j)\right]\times\bar{\bf P}_j\,,
 \nonumber\\
\end{eqnarray}
where we have ignored matter effects. Here $\mu\sim\sqrt2 G_{\rm F}
n_\nu$ is a measure of the neutrino-neutrino interaction energy in
the dense neutrino gas.

\begin{figure}[b]
\includegraphics[height=0.5\columnwidth]{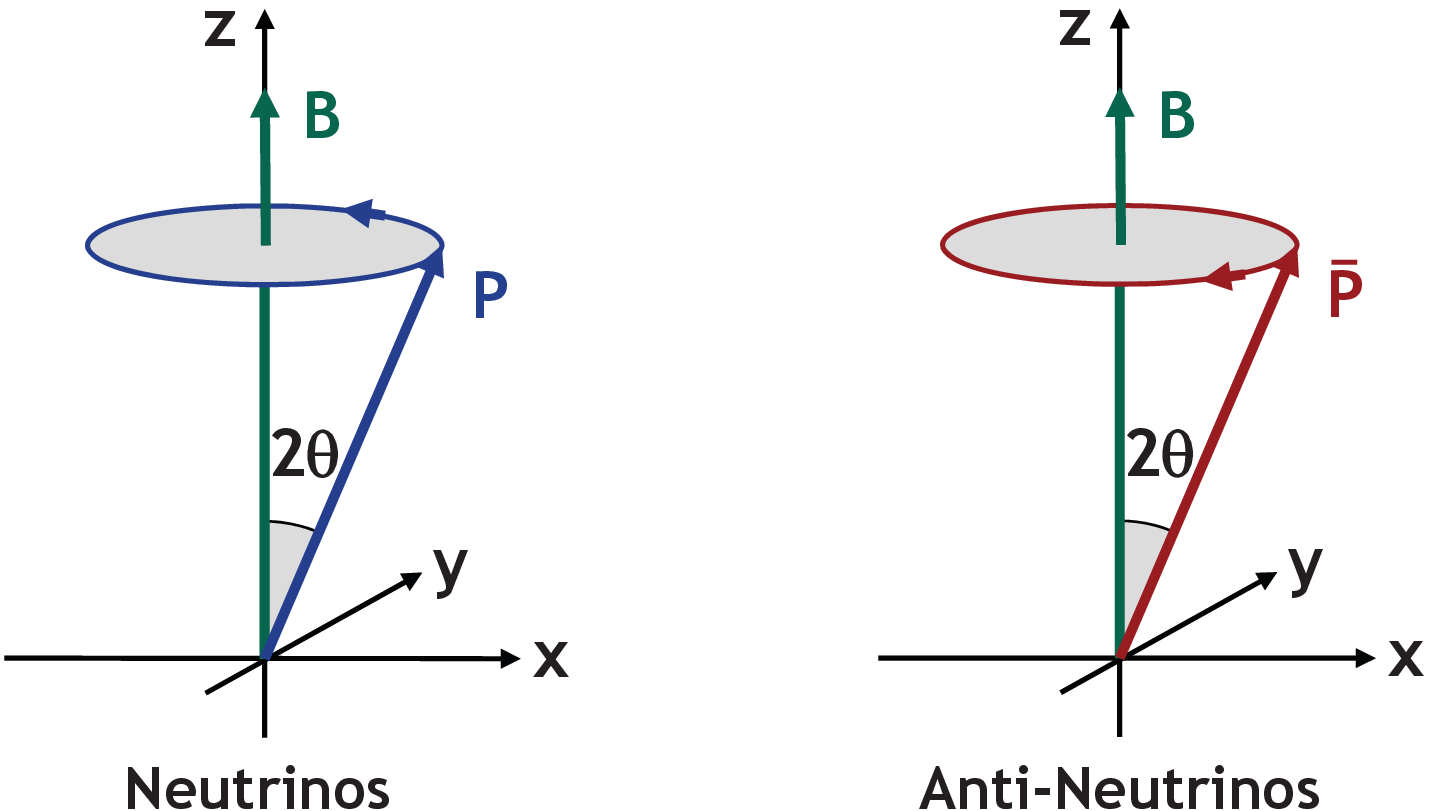}
\vskip15pt
\includegraphics[height=0.5\columnwidth]{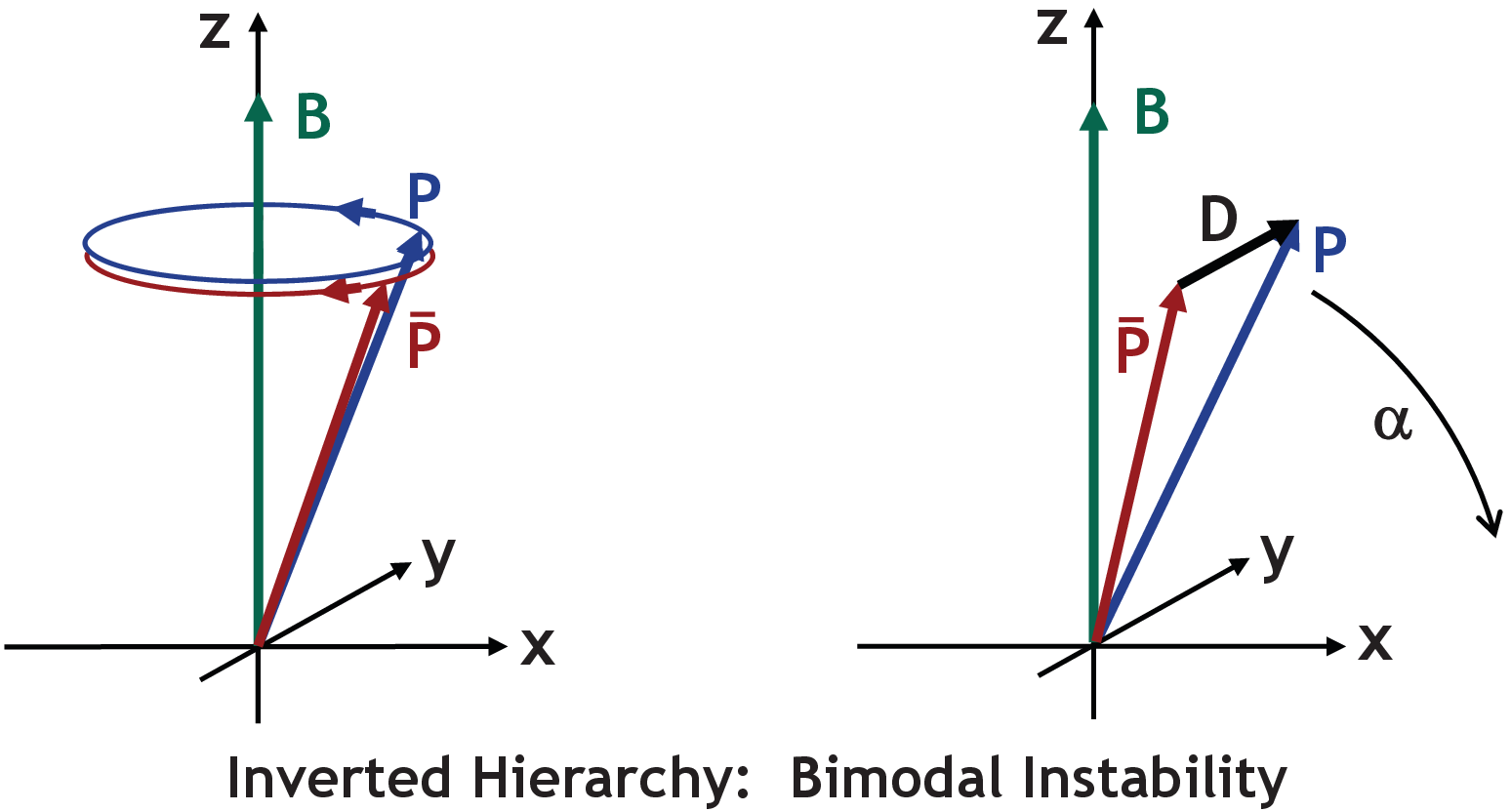}
\vskip15pt
\includegraphics[height=0.5\columnwidth]{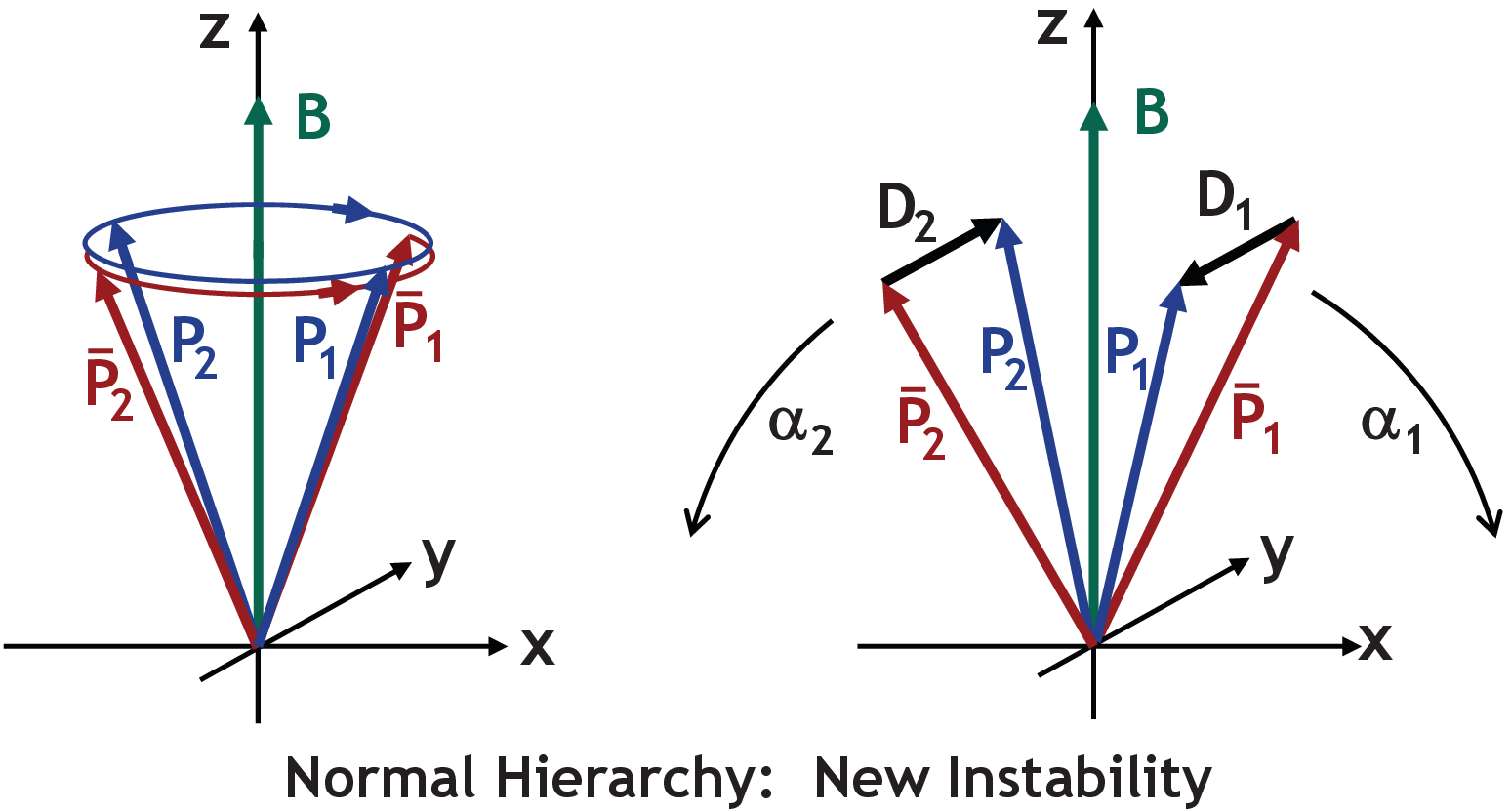}
\caption{Visualization of flavor oscillations and the bimodal and
new instability as described in the text.
\label{fig:precession}}
\end{figure}

For $\mu=0$ the evolution consists of a precession around the mass
direction ${\bf B}$ in flavor space in opposite directions for
neutrinos and anti-neutrinos (top-panel of
Fig.~\ref{fig:precession}). If neutrinos are prepared in flavor
eigenstates the initial polarization vectors are tilted relative to
${\bf B}$ by twice the vacuum mixing angle. We choose the tilting
direction to be in the $x$--$z$ plane in flavor space.

As a next step, we assume an isotropic ensemble and, following
established practice, sum over all polarization vectors and drop the
$1-{\vec v}_i\cdot{\vec v}_j$ factor. Of course, this is the crucial
step of assuming that solutions of the EoMs will be isotropic because
the system was prepared in such a state. In addition, we assume equal
energies for all particles. Then the EoMs are simply $\dot{\bf
P}=(\omega{\bf B}+\mu {\bf D})\times{\bf P}$, where ${\bf D}={\bf
P}-\bar{\bf P}$, and negative $\omega$ for anti-neutrinos. The EoMs
become even simpler when we introduce the variable ${\bf Q}={\bf
P}+\bar{\bf P}-(\omega/\mu){\bf B}$ so that $\dot{\bf Q}=\mu{\bf
D}\times{\bf Q}$ and $\dot{\bf D}=\omega{\bf B}\times{\bf Q}$. The
EoM for ${\bf Q}$ involves a cross product so that the length of
${\bf Q}$ is conserved and one can show that its dynamics is that of
a gyroscopic pendulum~\cite{Hannestad:2006nj}. For exact
particle-antiparticle symmetry the solution is even simpler. Taking
initially ${\bf P}=\bar{\bf P}$ and thus ${\bf D}=0$, the motion of
${\bf Q}$ is that of a plane pendulum, representing the generic case
of the bimodal instability.

In a pictorial explanation of this effect one begins with all
polarization vectors initially aligned with the flavor direction. The
flavor direction is taken to be nearly aligned with ${\bf B}$ up to
the mismatch caused by the small mixing-angle, i.e., we assume IH.
The motion begins as pure vacuum oscillations (middle panel of
Fig.~\ref{fig:precession}). The opposite direction of precession
drives ${\bf P}$ and $\bar{\bf P}$ apart and after a short while a
non-zero vector ${\bf D}$ has developed. Assuming that $\mu\gg\omega$
and when ${\bf D}$ has become sufficiently long, the further
evolution will be dominated by $\mu{\bf D}$ which points in the $y$
direction. So now both ${\bf P}$ and $\bar{\bf P}$ precess around
${\bf D}$ in a direction away from ${\bf B}$. This is the essence of
the bimodal instability. Of course, one needs to go through the
equations in detail to recognize that the full motion is that of an
inverted pendulum.

One way of switching the neutrino mass hierarchy in these equations
is $\Delta m^2\to-\Delta m^2$ or $\omega\to-\omega$. The
polarization vectors now precess in the opposite directions around
${\bf B}$ so that the generated ${\bf D}$ points in the opposite
direction. Now the precession around ${\bf D}$ has opposite sign and
both ${\bf P}$ and $\bar{\bf P}$ move toward ${\bf B}$, not away
from it. If the initial misalignment with ${\bf B}$ was small, the
resulting motion is a small-amplitude harmonic swinging around the
${\bf B}$ direction.

Next we consider our two-beam example with opposite velocity vectors.
The EoMs are now explicitly
\begin{eqnarray}\label{eq:eom2}
 \partial_t{\bf P}_1&=&\left(+\omega{\bf B}+\mu{\bf D}_2\right)\times{\bf P}_1\,,
 \nonumber\\
 \partial_t\bar{\bf P}_1&=&\left(-\omega{\bf B}+\mu{\bf D}_2\right)\times\bar{\bf P}_1\,,
 \nonumber\\
 \partial_t{\bf P}_2&=&\left(+\omega{\bf B}+\mu{\bf D}_1\right)\times{\bf P}_2\,,
 \nonumber\\
 \partial_t\bar{\bf P}_2&=&\left(-\omega{\bf B}+\mu{\bf D}_1\right)\times\bar{\bf P}_2\,,
\end{eqnarray}
where ${\bf D}_i={\bf P}_i-\bar{\bf P}_i$ with $i=1$ or 2. Assuming
that the 1 and 2 modes evolve identically, we are back to the
previous case of an isotropic system, except that here we have a
reflection symmetry between 1 and 2.

However, the two-beam system supports another class of solutions that
obtain if we prepare the 1 and 2 modes initially tilted in opposite
directions. We consider NH so that the initial vacuum precession is
opposite to the previously unstable case. From the third panel in
Fig.~\ref{fig:precession} one easily gleans the resulting directions
of ${\bf D}_1$ and ${\bf D}_2$. The key point is that mode 1 feels
the refractive effect caused by mode 2, and the other way round.
Therefore, the precession around the respective ${\bf D}_{1,2}$
vectors leads to motions away from ${\bf B}$ and subsequently to a
pendulum with the 1 and 2 modes swinging in opposite directions. So
this configuration is unstable in NH.

Switching the hierarchy back to IH reverses the directions of the
${\bf D}_{1,2}$ vectors and results in the familiar small-amplitude
oscillation around the ${\bf B}$ direction. The anti-symmetric mode
is unstable in the opposite hierarchy case compared with the bimodal
instability.

Of course, if the system is prepared {\em precisely} such that
initially all polarization vectors are the same, then only the
traditional bimodal instability can occur. The smallest deviation
from this exact initial symmetry causes an admixture of the
anti-symmetric mode. The exponentially growing mode always wins.
Therefore, any realistic initial condition provides a seed for a
run-away solution in both hierarchies.

\section{Coupled Oscillators}
\label{sec:oscillators}

We now turn to a more formal analysis of the two-beam example. To
this end we write the EoMs somewhat more symmetrically by
introducing the sum vectors ${\bf S}_i={\bf P}_i+\bar{\bf P}_i$ with
$i=1$ or~2 and find
\begin{eqnarray}\label{eq:eom4}
 \dot{\bf S}_1&=&\mu{\bf D}_2\times{\bf S}_1+\omega{\bf B}\times{\bf D}_1\,,
 \nonumber\\
 \dot{\bf S}_2&=&\mu{\bf D}_1\times{\bf S}_2+\omega{\bf B}\times{\bf D}_2\,,
 \nonumber\\
 \dot{\bf D}_1&=&\omega{\bf B}\times{\bf S}_1+\mu{\bf D}_2\times{\bf D}_1\,,
 \nonumber\\
 \dot{\bf D}_2&=&\omega{\bf B}\times{\bf S}_2+\mu{\bf D}_1\times{\bf D}_2\,.
\end{eqnarray}
We simplify further as in the previous section and assume exact
particle-antiparticle symmetry. We also assume that all polarization
vectors are initially in the $x$--$z$ plane and that the
corresponding particle and anti-particle vectors are prepared
identically. The system then evolves symmetrically such that the
${\bf S}_i$ vectors move in the $x$--$z$ plane, whereas the ${\bf
D}_i$ vectors are oriented along the $y$ direction. Therefore, the
terms ${\bf D}_1\times{\bf D_2}$ drop out.

In analogy to the previous section, we can now easily identify two
solutions that are symmetric or anti-symmetric under
$1\leftrightarrow2$. In the symmetric case, all polarization vectors
are initially prepared identically. Then ${\bf D}_2={\bf D}_1$ and
we find two pairs of decoupled equations. Likewise, if the $1$ and
$2$ modes are prepared with opposite $x$ components, we have ${\bf
D}_2=-{\bf D}_1$. Therefore, the two eigenmodes correspond to
equations of the form
\begin{eqnarray}\label{eq:eom5}
 \dot{\bf S}&=&\pm\mu{\bf D}\times{\bf S}+\omega{\bf B}\times{\bf D}\,,
 \nonumber\\
 \dot{\bf D}&=&\omega{\bf B}\times{\bf S}\,.
\end{eqnarray}
Changing the sign of $\mu$, like changing the sign of $\omega$,
corresponds to a change in hierarchy. Therefore, the solutions of
one case in one hierarchy is identical to the solution of the other
case with opposite hierarchy.

These pure symmetric or anti-symmetric solutions can each be brought
to the form of a pendulum equation by introducing the vector ${\bf
Q}={\bf S}\mp (\omega/\mu)\,{\bf B}$. The EoMs are then $\dot{\bf
Q}=\pm\mu{\bf D}\times{\bf Q}$ and $\dot{\bf D}=\omega{\bf
B}\times{\bf Q}$. However, the vector ${\bf Q}$ is defined
differently for the two cases. Still, if we go to the limit
$\mu\gg\omega$ we can approximately equal ${\bf S}$ with ${\bf Q}$
and the motion of each ${\bf S}_i$ is approximately a circle in the
$x$--$z$ plane with its center at the origin. We thus seek solutions
of the form
\begin{equation}
{\bf S}_i(t)=S_i\begin{pmatrix}\sin\alpha_i(t)\\ 0\\ \cos\alpha_i(t)\end{pmatrix}
\quad\hbox{and}\quad
{\bf D}_i(t)=\begin{pmatrix}0\\ D_i(t)\\0\end{pmatrix}
\end{equation}
leading to $\dot\alpha_1=\mu D_2$, $\dot\alpha_2=\mu D_1$, and $\dot
D_i=\omega S_i\sin\alpha_i$. Taking the second derivative then leads
to the oscillator equations $\ddot\alpha_1=\omega\mu
S_2\,\sin\alpha_2$ and $\ddot\alpha_2=\omega\mu S_1\,\sin\alpha_1$.
With $S=S_1=S_2$ and $\kappa^2=\omega\mu S$ we finally have
\begin{eqnarray}\label{eq:eom3}
 \ddot\alpha_1&=&\kappa^2\,\sin\alpha_2\,,
 \nonumber\\
 \ddot\alpha_2&=&\kappa^2\,\sin\alpha_1\,.
\end{eqnarray}
Our simplified system is therefore equivalent to two maximally mixed
anharmonic oscillators.

The eigenmodes of this system consist of a symmetric solution with
$\alpha_1(t)=\alpha_2(t)$ and an anti-symmetric one with
$\alpha_1(t)=-\alpha_2(t)$. In the former case, with
$\alpha=\alpha_1=\alpha_2$, the EoMs reduce to the single equation
$\ddot\alpha=\kappa^2\sin\alpha$. We always assume that initially
${\bf D}_i=0$, i.e., the particle and anti-particle modes begin in
the same flavor state so that $\dot\alpha(0)=0$. If the initial
angle is small, $\alpha(0)\ll 1$, the symmetric solution corresponds
to an inverted pendulum, i.e., an anharmonic oscillator beginning
near the maximum of the potential. Switching the hierarchy amounts
to $\omega\to-\omega$ and thus $\kappa^2\to-\kappa^2$. In this case,
the symmetric solution corresponds to a harmonic oscillator with a
small initial amplitude.

For the anti-symmetric solution we write $\alpha=\alpha_1=-\alpha_2$
so that the EoMs reduce to $\ddot\alpha=-\kappa^2\sin\alpha$.
Therefore, the original case of IH now corresponds to the
small-amplitude harmonic oscillator. Switching the hierarchy changes
$\kappa^2\to-\kappa^2$ and we are back to the inverted pendulum. In
other words, switching the hierarchy and switching between the
symmetric and anti-symmetric solution each causes a sign change in
$\ddot\alpha=\pm\kappa^2\sin\alpha$.

We may prepare the system initially such that both modes have
exactly equal particle densities in exact flavor eigenstates. In
this case $\alpha_1(0)=\alpha_2(0)$ and only the symmetric mode is
excited. This is the traditional case where the solution was assumed
to inherit the symmetry of the initial condition and small
deviations from that symmetry were assumed to remain small.

However, the latter assumption is unjustified in the case of NH
where the traditional solution is stable, whereas the anti-symmetric
solution is unstable. The smallest deviation of the initial state
from perfect symmetry provides a seed for the exponentially growing
solution which then eventually dominates.

\begin{figure}
\includegraphics[width=0.85\columnwidth]{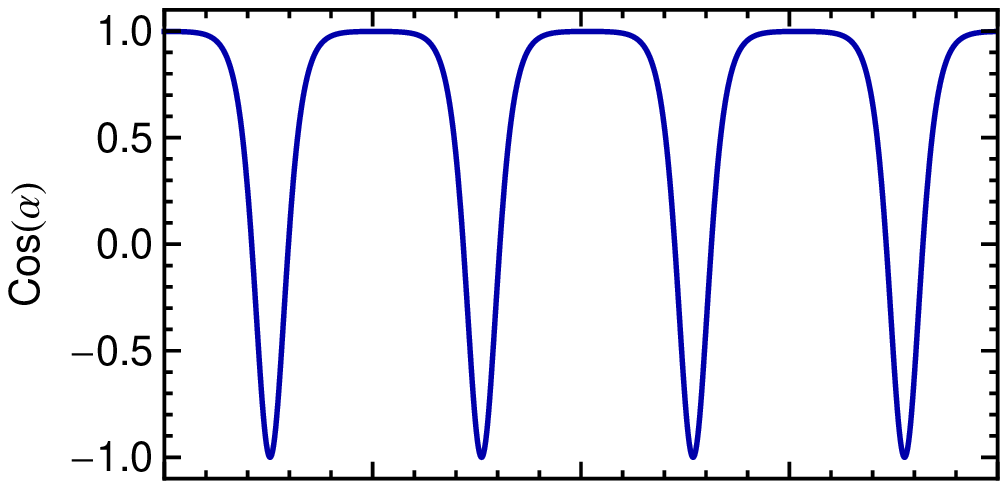}
\vskip3pt
\includegraphics[width=0.85\columnwidth]{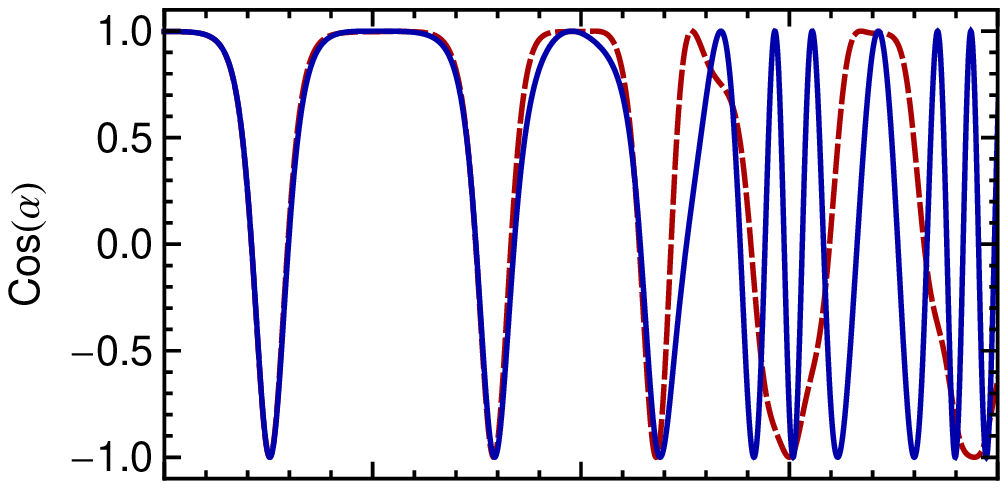}
\vskip3pt
\includegraphics[width=0.85\columnwidth]{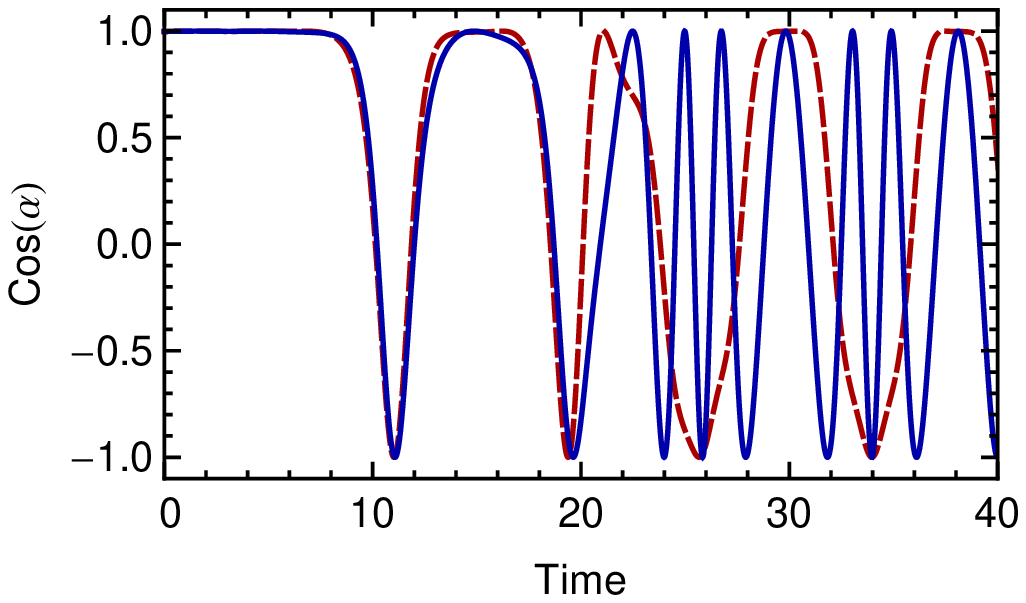}
\caption{Solution for $\alpha_1$ (blue solid line) and
$\alpha_2$ (red dashed line) with $\kappa=1$ and
$\alpha_1(0)=\alpha_2(0)=0.05$.
{\em Top:} Bimodal solution in IH.
{\em Middle:} Same with 1\% difference in $\kappa^2$ in the two EoMs.
{\em Bottom:} Same in NH, showing the new instability.
\label{fig:evolution}}
\end{figure}

We illustrate this behavior with a simple example in
Fig.~\ref{fig:evolution}. We use $\kappa=1$ and the initial angles
$\alpha_1(0)=\alpha_2(0)=0.05$. In IH we then find the usual bimodal
solution with pendulumlike periodic flavor conversion (top panel). To
disturb the perfect initial symmetry we can either make the initial
angles slightly different or assume slightly different particle
fluxes in the two beams. We choose the latter approach, implying that
$S_1$ is slightly different from $S_2$, i.e., $\kappa^2$ in the two
equations is a bit different. We use a 1\% difference. In this case
the bimodal solution at first proceeds as before, but it is now
coupled to the harmonic oscillator of the anti-symmetric solution.
Therefore, after a few pendulum dips, $\alpha_1(t)$ and $\alpha_2(t)$
get out of step, resulting in chaotic behavior.

Switching the hierarchy (bottom panel) would provide the traditional
stable solution if the initial conditions were perfectly symmetric.
Implementing our small mismatch has at first no effect, but the
unstable admixture of the anti-symmetric mode quickly grows enough to
take over and again we get pendulumlike behavior that also turns
chaotic.

Even for exact initial symmetry, numerical noise is enough to
trigger the run-away solution in NH, at least when using Mathematica
to solve the equations with standard settings. In our example, the
first dip triggered by numerical noise was roughly at $t=45$.

Our example was constructed to be perfectly symmetric between
particles and antiparticles, restricting the motion to the $x$--$z$
plane and we have assumed $\mu\gg\omega$. Numerical examples solving
the full EoMs of Eq.~(\ref{eq:eom4}) in various cases show analogous
effects. In other words, if we relax further symmetry assumptions
does not prevent the new instability, but also does not introduce
yet other qualitatively new effects. From a linearized stability
analysis of our general two-beam system we indeed find exactly two
instabilities. Of course, in the general case the instabilities only
arise for certain ranges of $\mu$ values.

\section{Conclusion}
\label{sec:conclusion}

We have constructed the simplest example that graphically
illustrates the recent insight that the multi-angle nature of
neutrino-neutrino refraction can lead to instabilities that
spontaneously break the symmetry of initial conditions. In
particular, we have found that the traditional flavor pendulum
actually occurs in both hierarchies if one allows angular modes to
evolve independently.

When the neutrino and anti-neutrino populations are not identical,
the more realistic case in situations of practical interest, the
system is stable at large densities. It remains to be seen if, when
slowly reducing the density, one can obtain ordered solutions and
spectral features such as spectral splits. Likewise, one can ask for
the existence of pure precession modes, their stability, and many
other properties.

Our simple example was meant as a proxy for the somewhat confusing
multi-azimuth angle (MAA) instability that occurs in axially
symmetric SN neutrino fluxes. It was found that this instability
occurs even when using only two discrete azimuth angles, i.e., for a
two-beam example. In the SN case, these two modes intersect at a
small angle, but have opposite directions in the transverse plane.
As the magnitude of the intersection angle only influences the
effective neutrino-neutrino interaction strength, our two-beam
example closely mimics this case. The formal stability analysis
reveals that there is only one additional instability, not one for
every discrete azimuth direction. The traditional bimodal
instability has no $\varphi$ dependence, the new instability a
dipole structure in $\varphi$.

Relaxing unjustified symmetry assumptions concerning the solutions
of the equations of collective flavor oscillations leads to new
effects. What this means in practice, for example in the context of
SN neutrino oscillations, remains to be explored. It is certain,
however, that one cannot trust any result that relies on symmetries
that may get spontaneously broken by the interacting neutrino
system.


\section*{Acknowledgments}

This work was partly supported by the Deutsche
Forschungsgemeinschaft under grant EXC-153 (Cluster of Excellence
``Origin and Structure of the universe'') and by the European Union
under grant PITN-GA-2011-289442 (FP7 Initial Training Network
``Invisibles''). D.~S. acknowledges support by the Funda\c{c}\~{a}o
para a Ci\^{e}ncia e Tecnologia (Portugal) under grant
SFRH/BD/66264/2009.


\end{document}